\documentclass[10pt,keeplastbox]{article}

\usepackage{SKConferencePaperV0003}

\let\OLDthebibliography\thebibliography
\renewcommand\thebibliography[1]{
  \OLDthebibliography{#1}
  \setlength{\parskip}{0pt}
  \setlength{\itemsep}{5pt plus 0.3ex}
}

\title{Spectral Statistics of Directed Networks with\\Random Link Model Transpose-Asymmetry}

\name{\vspace{-0pt}\parbox{\linewidth}{\centering
Stephen Kruzick and Jos\'{e} M. F. Moura\\~\\
\textnormal{
Carnegie Mellon University, Department of Electrical Engineering\\
5000 Forbes Avenue, Pittsburgh, Pennsylvania 15213}\vspace{-5pt}}
\thanks{Stephen~Kruzick (skruzick@andrew.cmu.edu) and Jos\'{e}~M.~F.~Moura (moura@andrew.cmu.edu) are with the Department of Electrical and Computer Engineering at Carnegie Mellon University in Pittsburgh, PA, USA.  This work was supported by NSF grant \#~CCF~1513936.}}
\address{}

\begin{document}
\maketitle

\setlength{\abovedisplayskip}{4pt}
\setlength{\belowdisplayskip}{4pt}
\setlength{\abovedisplayshortskip}{2pt}
\setlength{\belowdisplayshortskip}{2pt}

\begin{spacing}{.99}
\begin{abstract}%\vspace{-1pt}
Stochastic network influences complicate graph filter design by producing uncertainty in network iteration matrix eigenvalues, the points at which the graph filter response is defined.  While joint statistics for the eigenvalues typically elude analysis, predictable spectral asymptotics can emerge for large scale networks.  Previously published works successfully analyze large-scale networks described by undirected graphs and directed graphs with transpose-symmetric distributions, focusing on consensus acceleration filter design for time-invariant networks as an application.  This work expands upon these results by enabling analysis of certain large-scale directed networks described by transpose-asymmetric distributions.  Specifically, efficiently computable spectral density approximations are possible for transpose-asymmetric percolation network models with node-transitive symmetry group and normal mean matrix.  Numerical simulations support the derived approximations and application to consensus filters.
\end{abstract}

\end{spacing}

\vspace{-3pt}
\section{Introduction}\label{Introduction}
\vspace{-2pt}
In order to handle modern data sources with relationships described by an underlying network structure \cite{ASan2}, graph signal processing techniques have been developed to analyze data supported by the nodes of a graph~\cite{DShu1}.  In graph signal processing, the shift operator $W$ is defined by some matrix that respects the graph structure, such as the adjacency matrix~\cite{ASan1} or Laplacian matrix~\cite{DShu1}.  Shift-invariant filters arise as polynomials $p\left(W\right)$ in this shift matrix, with filter response defined at the eigenvalues $\{\lambda_i(W)\}$ of $W$~\cite{ASan3}.  Therefore, for scenarios described by random graphs, the associated shift matrix and its eigenvalues become random as they depend on the graph structure, significantly complicating filter design problems.  However, for networks with many nodes emergent predictability can arise from the large-scale nature of the problem that can be turned to advantage. This phenomenon comes in the form of limiting theorems from random matrix theory.  Although the joint distribution of these eigenvalues is typically not tractable, the empirical distribution built from the random eigenvalues of the random matrix sometimes has a deterministic limit as the size of the matrix grows for suitable models~\cite{RCou1}.  The Wigner semicircular law~\cite{EWig1}, the Marchenko-Pastur law~\cite{VMar1}, and the Girko circular law~\cite{VGir1} represent three well-known examples of such limiting behavior for large-scale random matrices.  Such deterministic approximations to the true empirical spectral distribution can provide useful information for graph filter design problems.  For example, consider distributed average consensus, the task of iteratively averaging all node data through only local network communications \cite{ROlfSab2}, which finds use in several applications~\cite{GCyb1,LXia2,SKar2,ROlfSab1}.  Graph filters applied at each node can accelerate consensus convergence~\cite{SSun1,ASan4,EKok1,EMon1,ALou1,SApe1,FGam1}, which can benefit from asymptotic spectral information\mbox{ for suitable random networks~\cite{SKru1,SKru2,SKru3,SKru4,SKru5,SKru6}.}

This paper focuses on computing deterministic approximations to the empirical spectral distribution of matrices that respect the structure of large-scale random directed networks, information that is useful for graph filter design optimization problems on random topologies when the distribution is known.  The work in this paper accommodates random network models where the link directions have different probabilities, leading to random matrix models that have different distribution from their transposes.  In contrast,~\cite{SKru6} has a similar purpose but focuses more closely on the filter design aspects and only examines directed networks in which both link directions have identical distribution.  Under the less restrictive conditions of this paper, additional analysis is necessary.

Section~\ref{NonSymmetricMean} introduces a theorem by Girko~\cite{VGir1} useful for describing the spectral asymptotics of random non-Hermitian matrices that arise from large-scale random directed networks with statistically independent links.  The section then proceeds to discuss practical numerical computation of an approximation to the empirical spectral distribution for random matrix distributions that are not transpose-symmetric.  Finally, it applies the observations to an example class of matrices with non-transpose-symmetric distribution.  Section~\ref{NumericalSimulations} demonstrates through numerical simulations both the approximation of the empirical spectral distribution through these methods and the application of this approximation to the consensus filter design optimization derived in~\cite{SKru6} for the case of directed networks with non-symmetric link probabilities.  Finally, Section~\ref{Conclusion} provides \mbox{concluding analysis.}

%\clearpage
%\input{Sections/FilterDesign_ComplexDN_Figures_NonSymmetricMean}
\section{Directed Networks:  \mbox{Non-symmetric Distributions}}\label{NonSymmetricMean}
\begin{figure*}[t]

\begin{floatrow}[1]
\ffigbox[\textwidth]{
\noindent\centering
\begin{minipage}{\textwidth}\centering
\begin{minipage}{.3\textwidth}
\ffigbox[\textwidth]{\includegraphics[width=\linewidth]{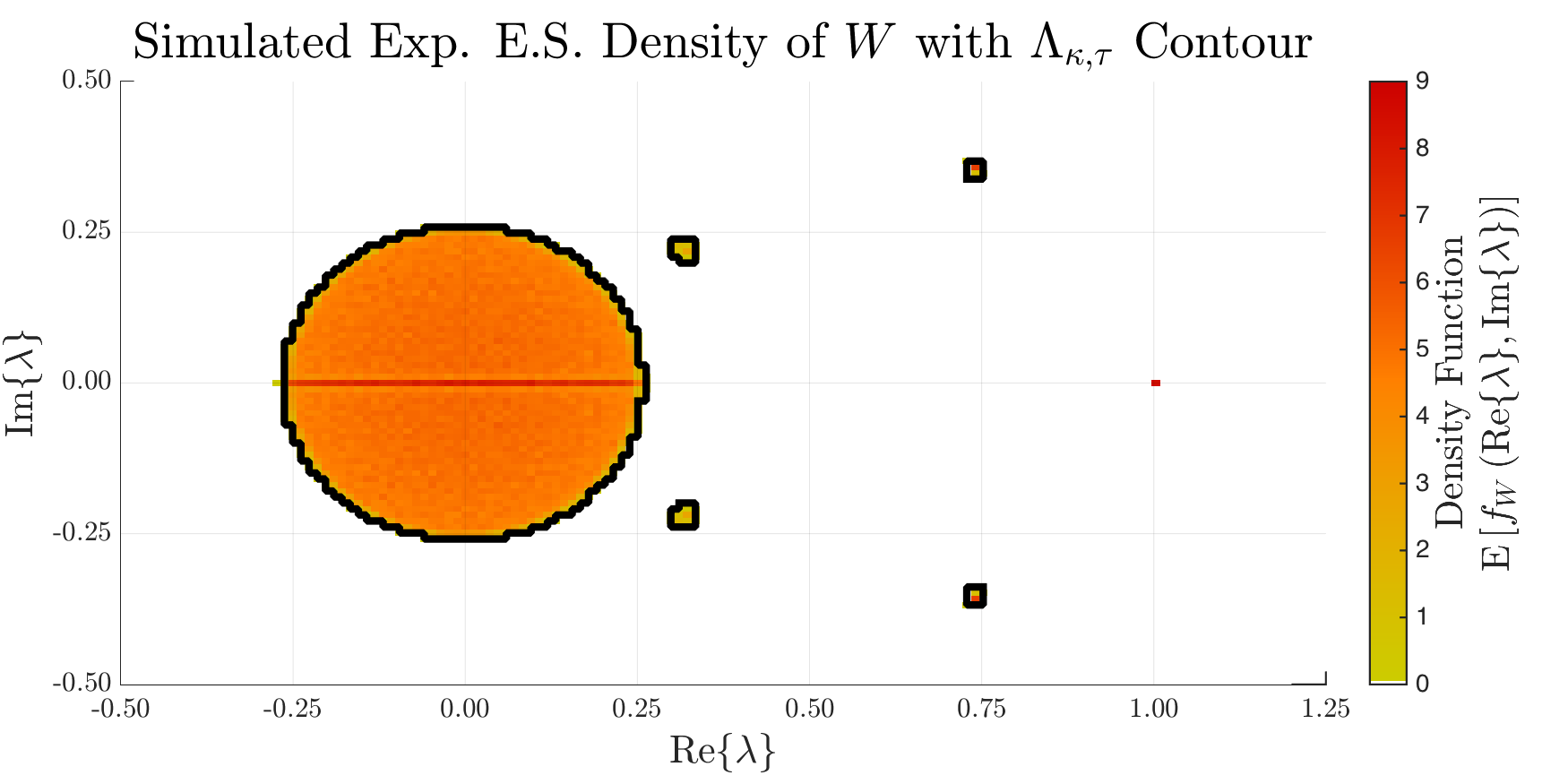}}{\caption{Expected empirical spectral density for the cyclic (non-symmetric) SBM with $M=5$ populations described in Sec.~\ref{NumericalSimulations} ($1000$ trials).  The contour shows the boundary of $\Lambda_{\kappa,\tau}$.}\label{Sim1a}}
\end{minipage}\hfill
\begin{minipage}{.3\textwidth}
\ffigbox[\textwidth]{\includegraphics[width=\linewidth]{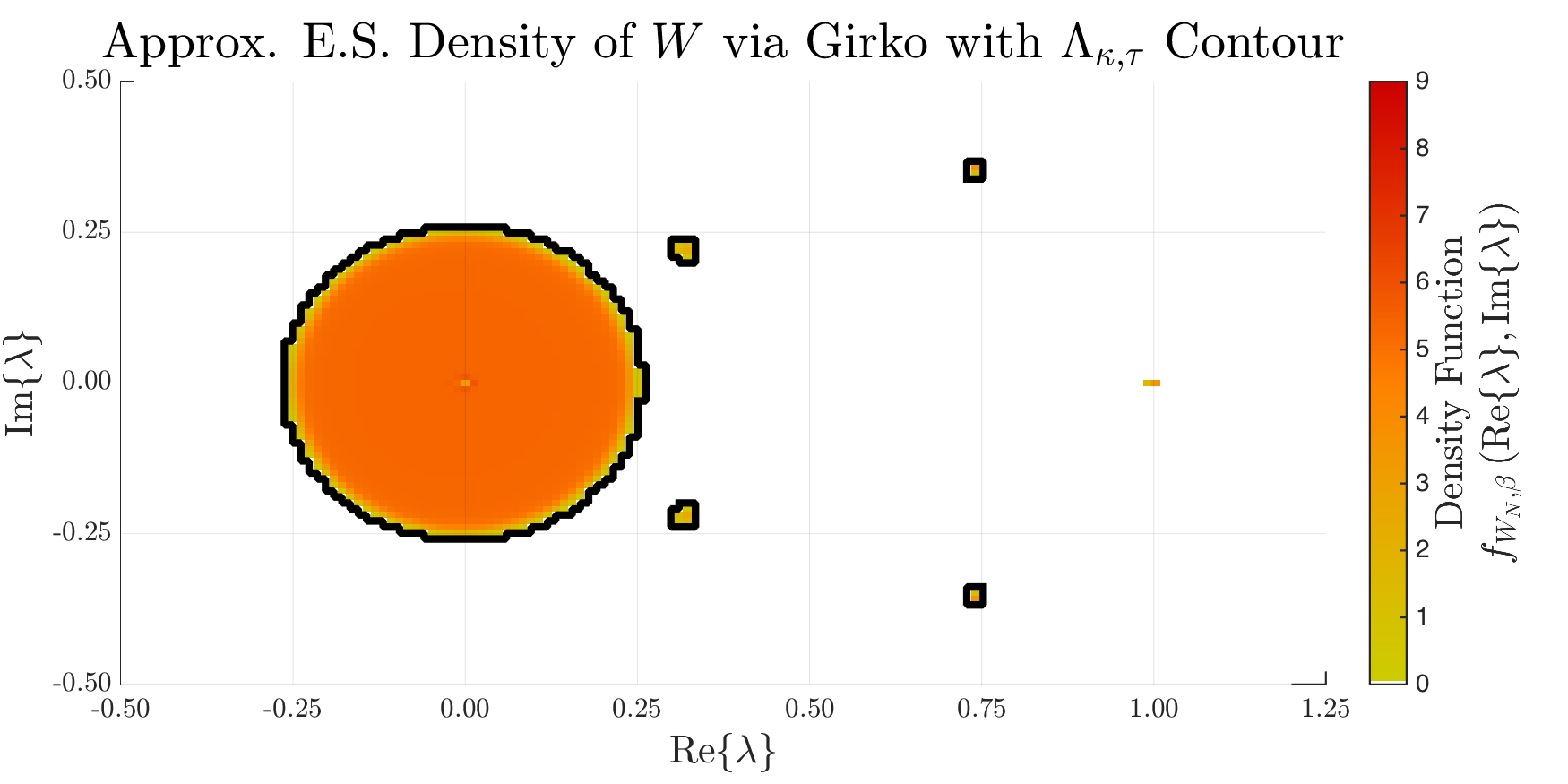}}{\caption{Approximate density computed via Girko's eq. as described in Sec.~\ref{NonSymmetricMean} for iteration matrix model from Fig.~\ref{Sim1a}.  The contour shows the boundary of $\Lambda_{\kappa,\tau}$ derived from $\widehat{f}_{W_N,\beta}$.}\label{Sim1b}}
\end{minipage}\hfill
\begin{minipage}{.3\textwidth}
\ffigbox[\textwidth]{\includegraphics[width=\linewidth]{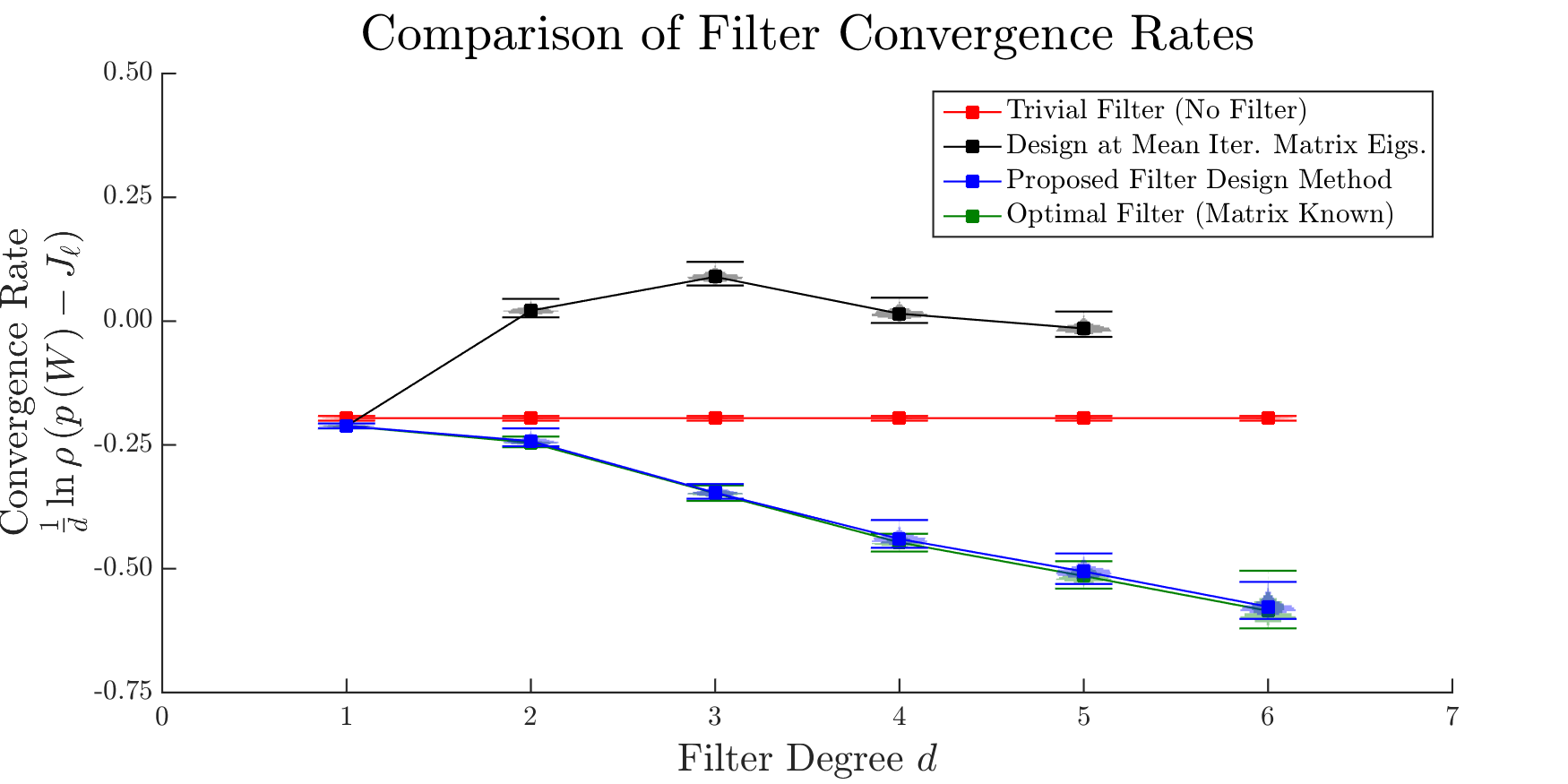}}{\caption{Consensus convergence rates (log scale, per degree) for network model from Fig.~\ref{Sim1a} for several filters of degrees $d=1,\ldots,6$.  Results averaged over $1000$ Monte-Carlo trials.}\label{Sim1c}}
\end{minipage}
\end{minipage}}{}
\vspace{5pt}
\end{floatrow}
\begin{floatrow}[1]
\ffigbox[\textwidth]{
\noindent\centering
\begin{minipage}{\textwidth}\centering
\noindent
\begin{subfloatrow}[4]
\ffigbox[.220\textwidth]{\includegraphics[width=\linewidth]{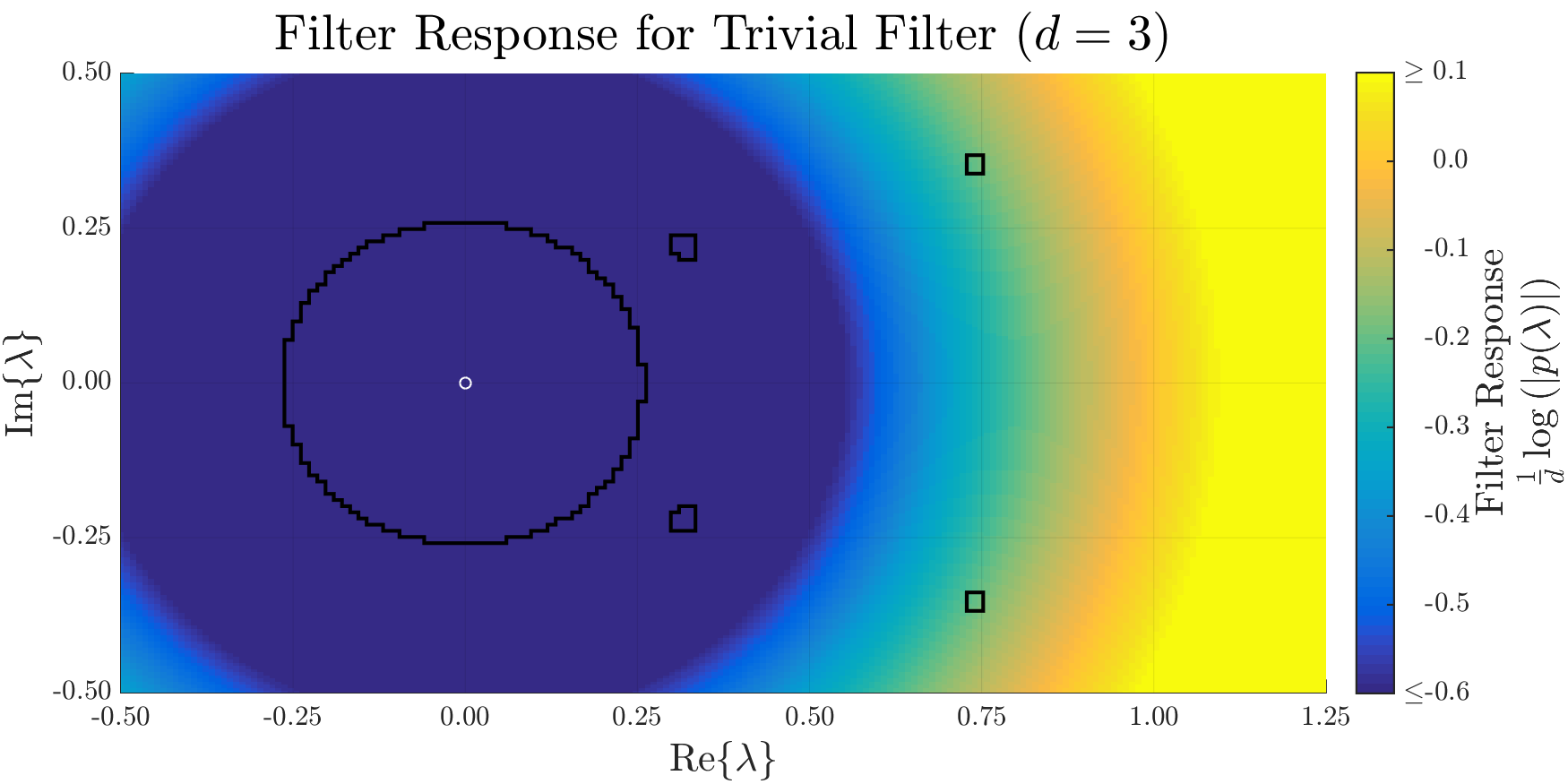}}{\caption{Proposed Filter ($d=3$)}\label{Sim1d1}}
\ffigbox[.220\textwidth]{\includegraphics[width=\linewidth]{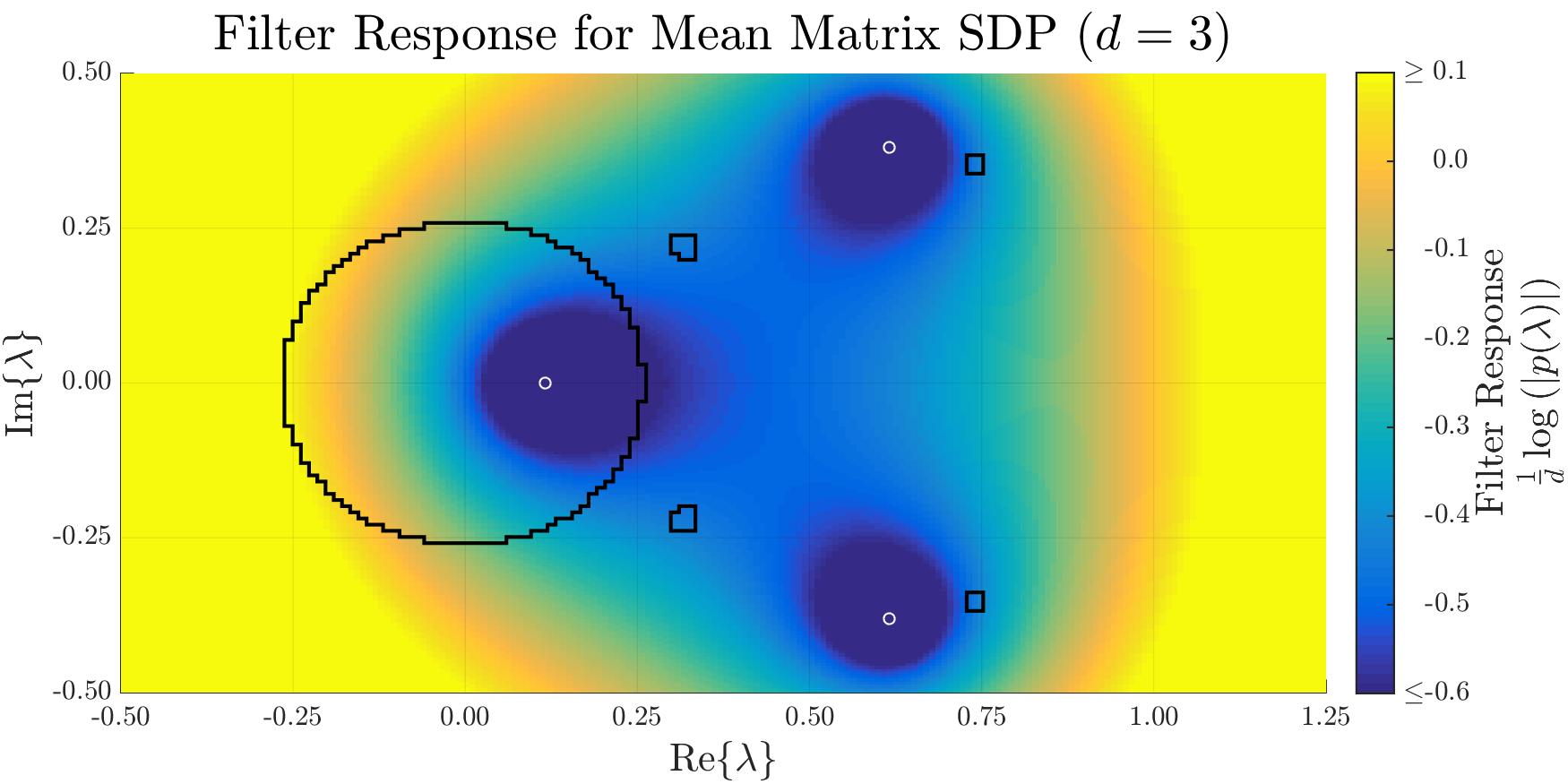}}{\caption{Ex. Optimal Filter ($d=3$)}\label{Sim1d2}}
\ffigbox[.220\textwidth]{\includegraphics[width=\linewidth]{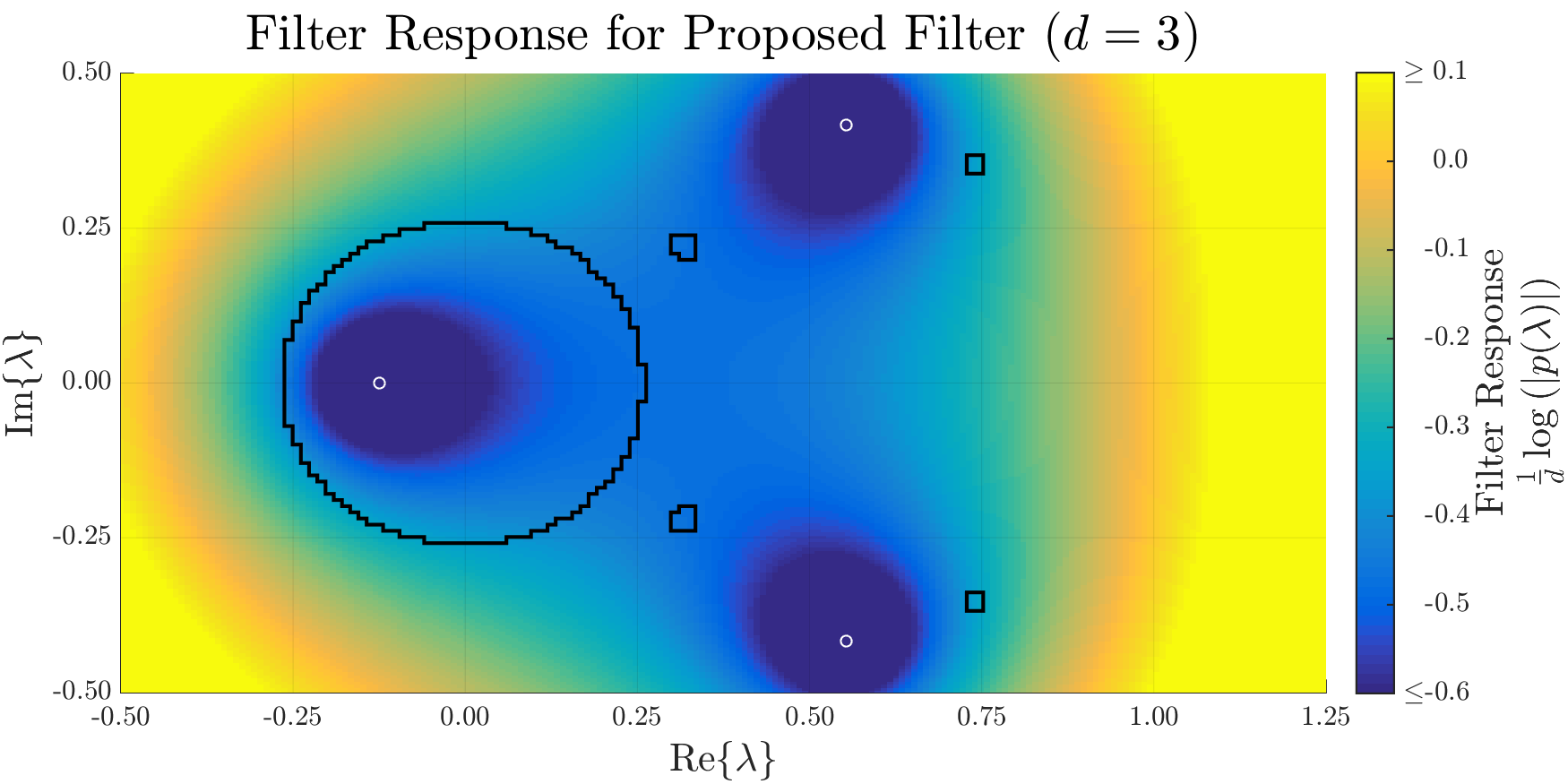}}{\caption{Proposed Filter ($d=3$)}\label{Sim1d3}}
\ffigbox[.220\textwidth]{\includegraphics[width=\linewidth]{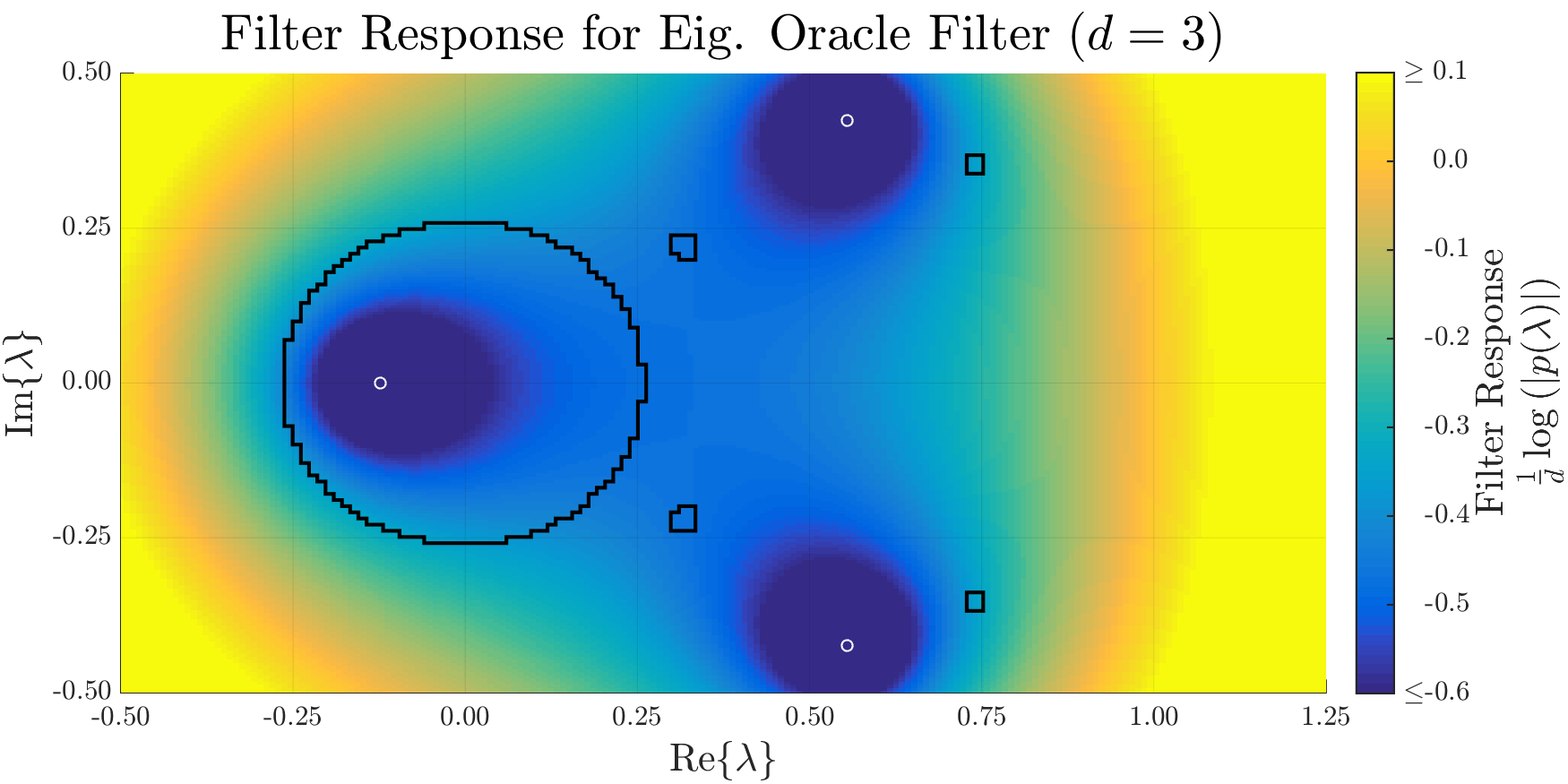}}{\caption{Ex. Optimal Filter ($d=3$)}\label{Sim1d4}}
\end{subfloatrow}

%\begin{minipage}{.475\textwidth}\centering
%%\ffigbox[\textwidth]{
%\begin{minipage}{\textwidth}\centering
%\begin{subfloatrow}[2]
%\ffigbox[.460\textwidth]{\includegraphics[width=\linewidth]{NumSim1D1}}{\caption{Trivial Filter ($d=3$)}\label{Sim1d1}}
%\ffigbox[.460\textwidth]{\includegraphics[width=\linewidth]{NumSim1D2}}{\caption{Mean Matrix SDP ($d=3$)}\label{Sim1d2}}
%\end{subfloatrow}\vspace{5pt}
%%\begin{subfloatrow}[2]
%%\ffigbox[.460\textwidth]{\includegraphics[width=\linewidth]{NumSim1D3}}{\caption{Proposed Filter ($d=3$)}\label{Sim1d3}}
%%\ffigbox[.460\textwidth]{\includegraphics[width=\linewidth]{NumSim1D4}}{\caption{Ex. Optimal Filter ($d=3$)}\label{Sim1d4}}
%%\end{subfloatrow}
%\end{minipage}
%%}{\caption{}\label{Sim1d}}
%\end{minipage}
%\hfill
%\begin{minipage}{.475\textwidth}\centering
%\ffigbox[\textwidth]{
%\begin{minipage}{\textwidth}\centering
%\begin{subfloatrow}[1]
%\ffigbox[.460\textwidth]{\includegraphics[width=\linewidth]{NumSim1D5}}{\caption{Trivial Filter ($d=6$)}\label{Sim1d5}}
%%\ffigbox[.460\textwidth]{\includegraphics[width=\linewidth]{NumSim1D6}}{\caption{Mean Matrix SDP ($d=6$)}\label{Sim1d6}}
%\end{subfloatrow}\vspace{5pt}
%\begin{subfloatrow}[2]
%\ffigbox[.460\textwidth]{\includegraphics[width=\linewidth]{NumSim1D7}}{\caption{Proposed Filter ($d=6$)}\label{Sim1d7}}
%\ffigbox[.460\textwidth]{\includegraphics[width=\linewidth]{NumSim1D8}}{\caption{Ex. Optimal Filter ($d=6$)}\label{Sim1d8}}
%\end{subfloatrow}
%\end{minipage}
%}{\caption{}\label{Sim1e}}
%\end{minipage}
\end{minipage}
}{\caption{Filter response magnitudes (log scale, per degree) for each filter type plotted in Fig.~\ref{Sim1c} (trivial filter, mean matrix SDP, proposed filter based on approx. density, and an example filter designed with known eigenvalues) are shown for degree $d=3$.  Locations of zeros are marked (white circles), and the boundary of the region $\Lambda_{\kappa,\tau}$ is shown (black contour).  Note that $E\left[W\right]$ has $K=6$ distinct eigs., so results are not shown for the mean matrix SDP method at $d=6$ as $K-1<6$.\vspace{-12pt}}\label{Sim1d}}

\end{floatrow}

\end{figure*}

\begin{figure*}[t]

\begin{floatrow}[1]
\ffigbox[\textwidth]{
\noindent
\begin{minipage}{\textwidth}\centering
\begin{minipage}{.3\textwidth}
\ffigbox[\textwidth]{\includegraphics[width=\linewidth]{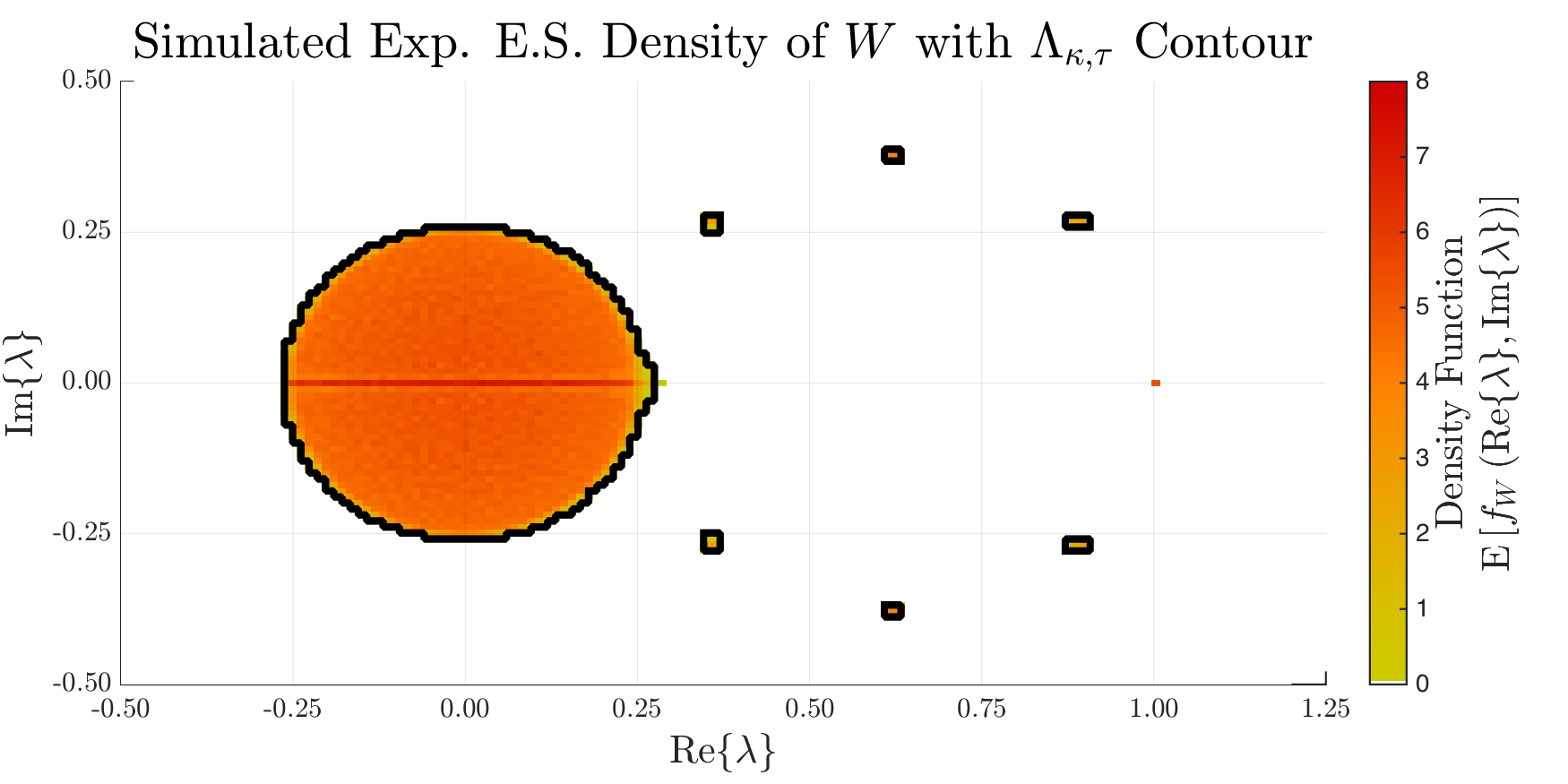}}{\caption{Expected empirical spectral density for the cyclic (non-symmetric) SBM with $M=8$ populations described in Sec.~\ref{NumericalSimulations} ($1000$ trials).  The contour shows the boundary of $\Lambda_{\kappa,\tau}$.}\label{Sim2a}}
\end{minipage}\hfill
\begin{minipage}{.3\textwidth}
\ffigbox[\textwidth]{\includegraphics[width=\linewidth]{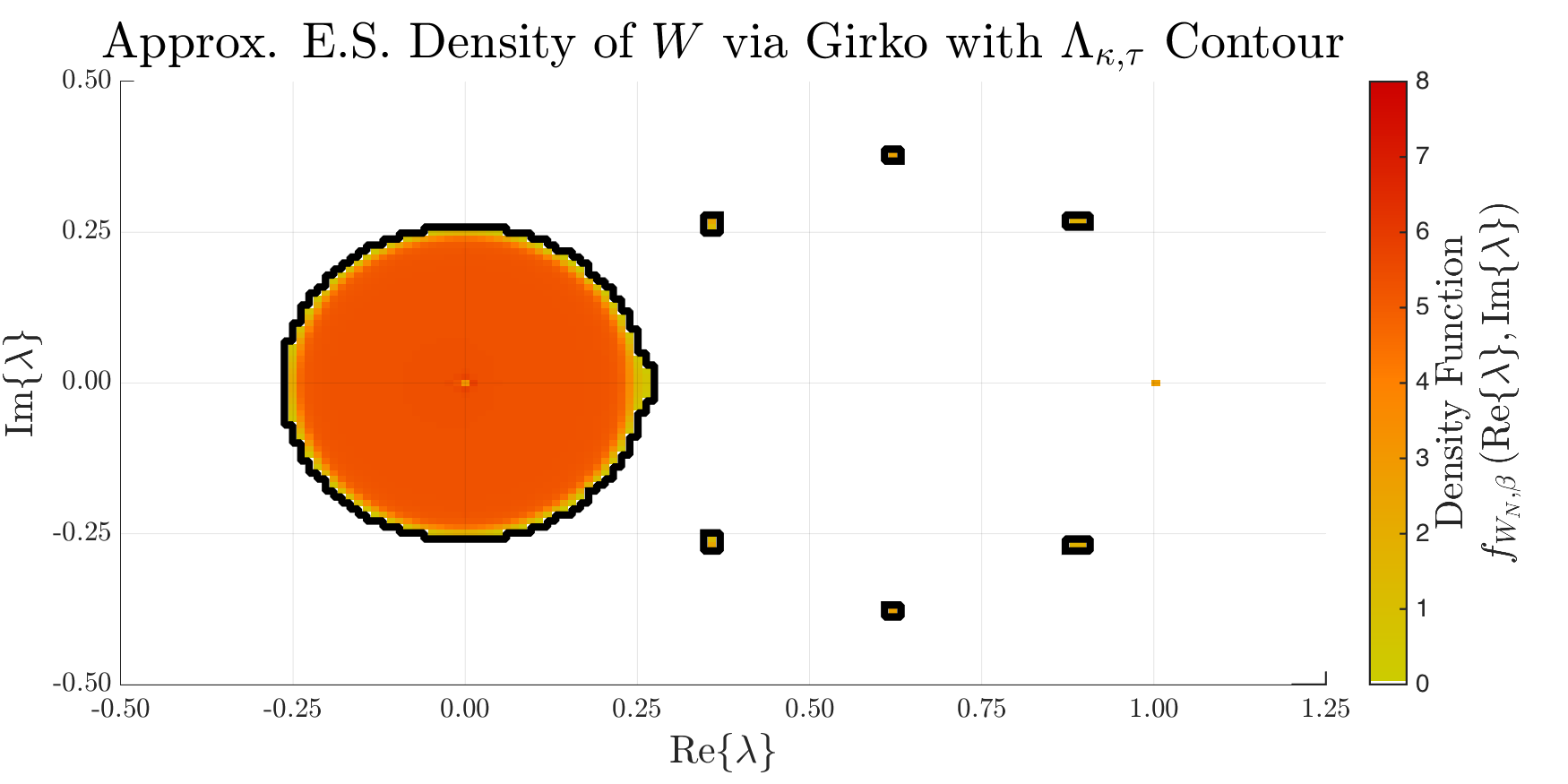}}{\caption{Approximate density computed via Girko's eq. as described in Sec.~\ref{NonSymmetricMean} for iteration matrix model from Fig.~\ref{Sim2a}.  The contour shows the boundary of $\Lambda_{\kappa,\tau}$ derived from $\widehat{f}_{W_N,\beta}$.}\label{Sim2b}}
\end{minipage}\hfill
\begin{minipage}{.3\textwidth}
\ffigbox[\textwidth]{\includegraphics[width=\linewidth]{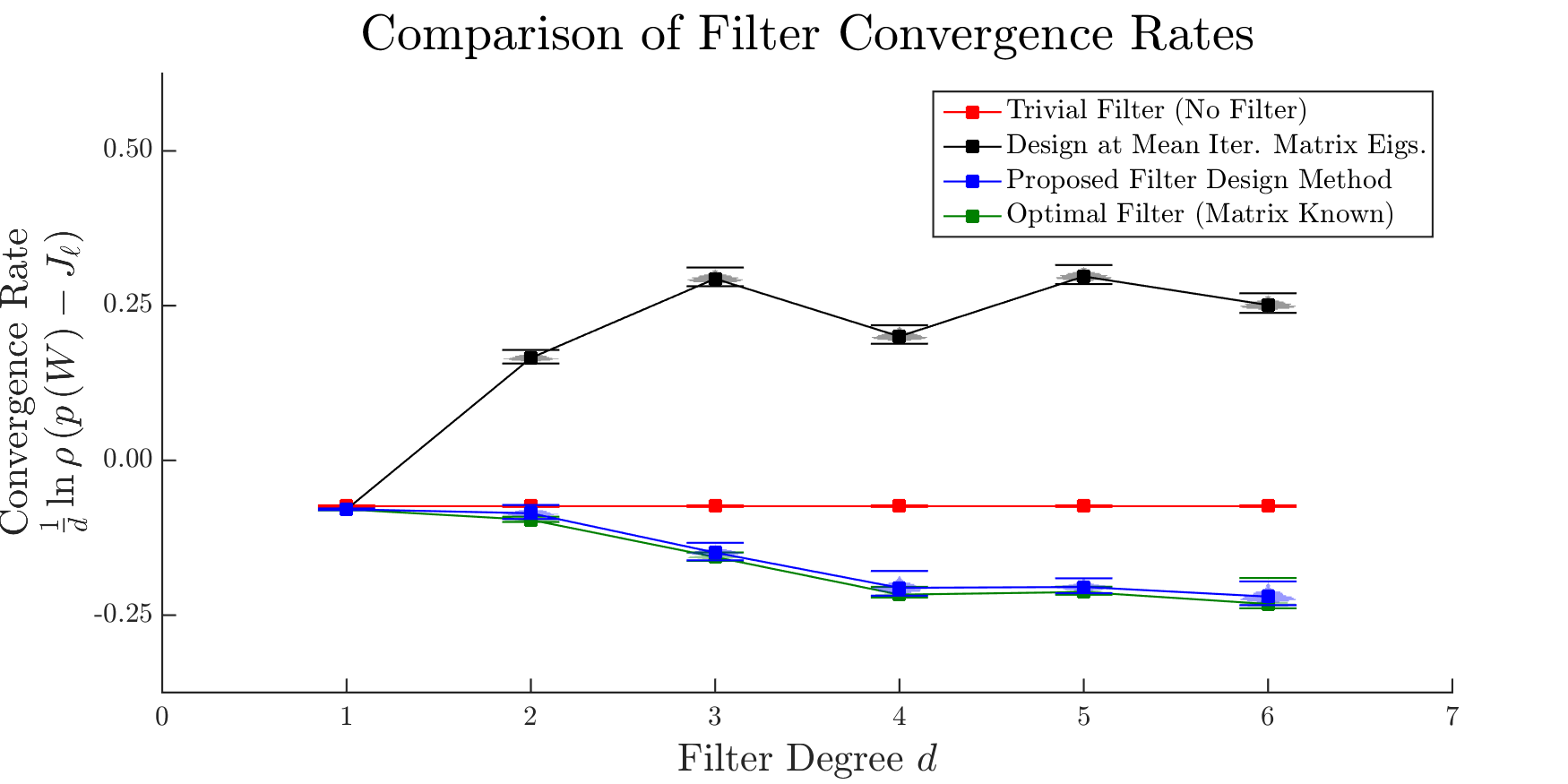}}{\caption{Consensus convergence rates (log scale, per degree) for network model from Fig.~\ref{Sim2a} for several filters of degrees $d=1,\ldots,6$.  Results averaged over $1000$ Monte-Carlo trials.}\label{Sim2c}}
\end{minipage}
\end{minipage}
}{}
\vspace{5pt}
\end{floatrow}

\begin{floatrow}[1]
\ffigbox[\textwidth]{
\noindent\hspace{-10pt}
\begin{minipage}{\textwidth}\centering
\noindent
\begin{subfloatrow}[4]
\ffigbox[.220\textwidth]{\includegraphics[width=\linewidth]{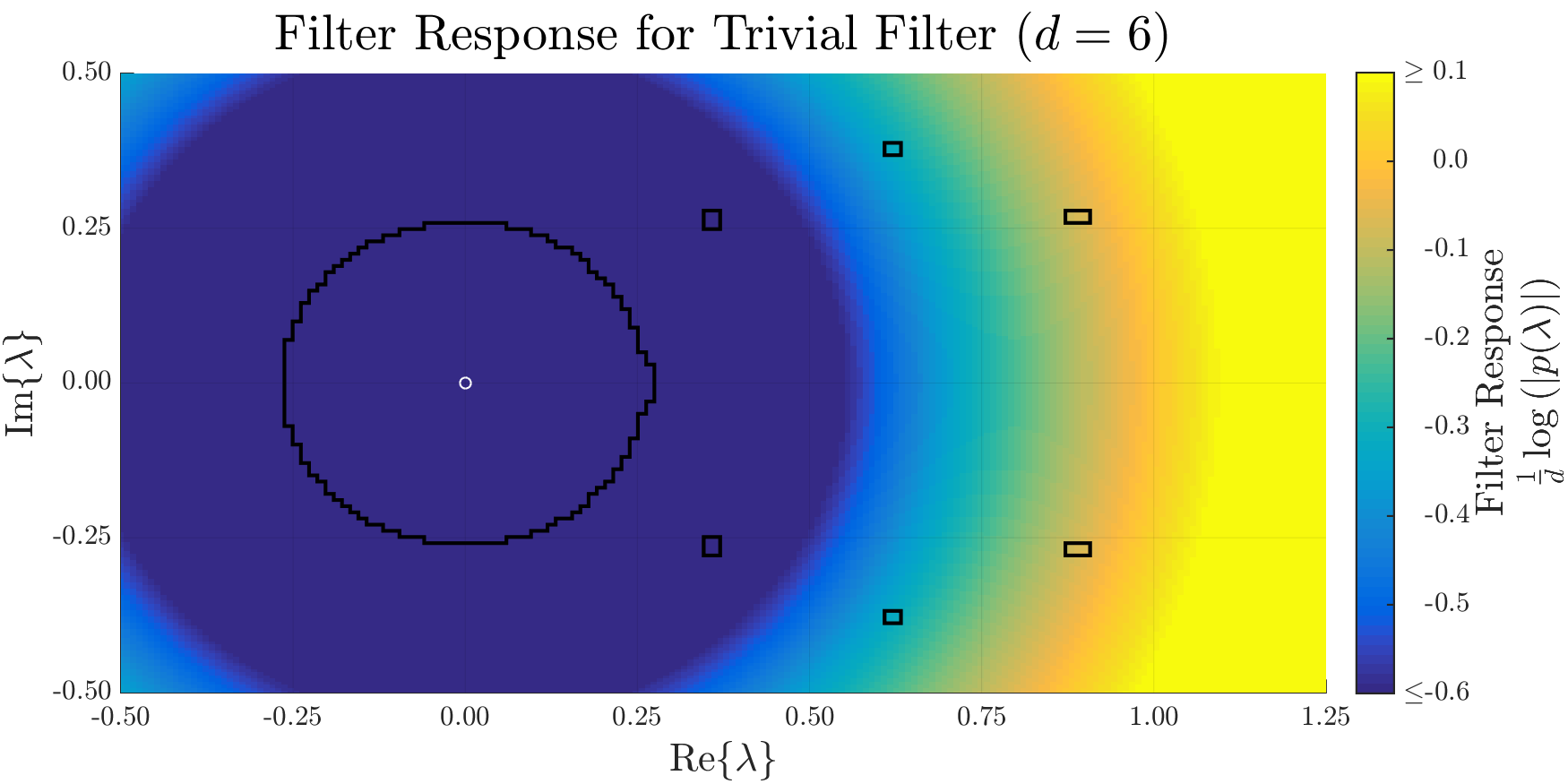}}{\caption{Proposed Filter ($d=6$)}\label{Sim2d1}}
\ffigbox[.220\textwidth]{\includegraphics[width=\linewidth]{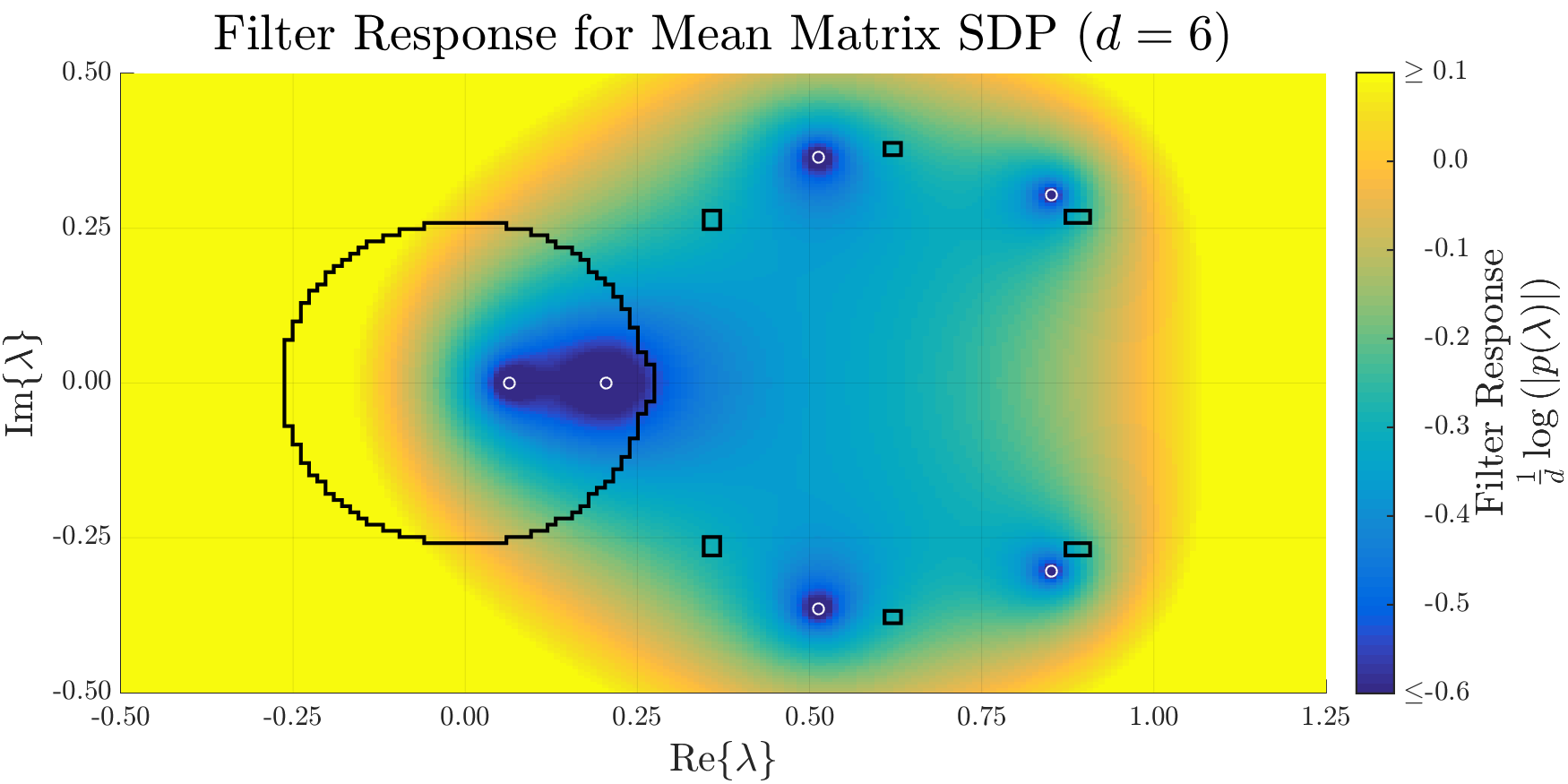}}{\caption{Ex. Optimal Filter ($d=6$)}\label{Sim2d2}}
\ffigbox[.220\textwidth]{\includegraphics[width=\linewidth]{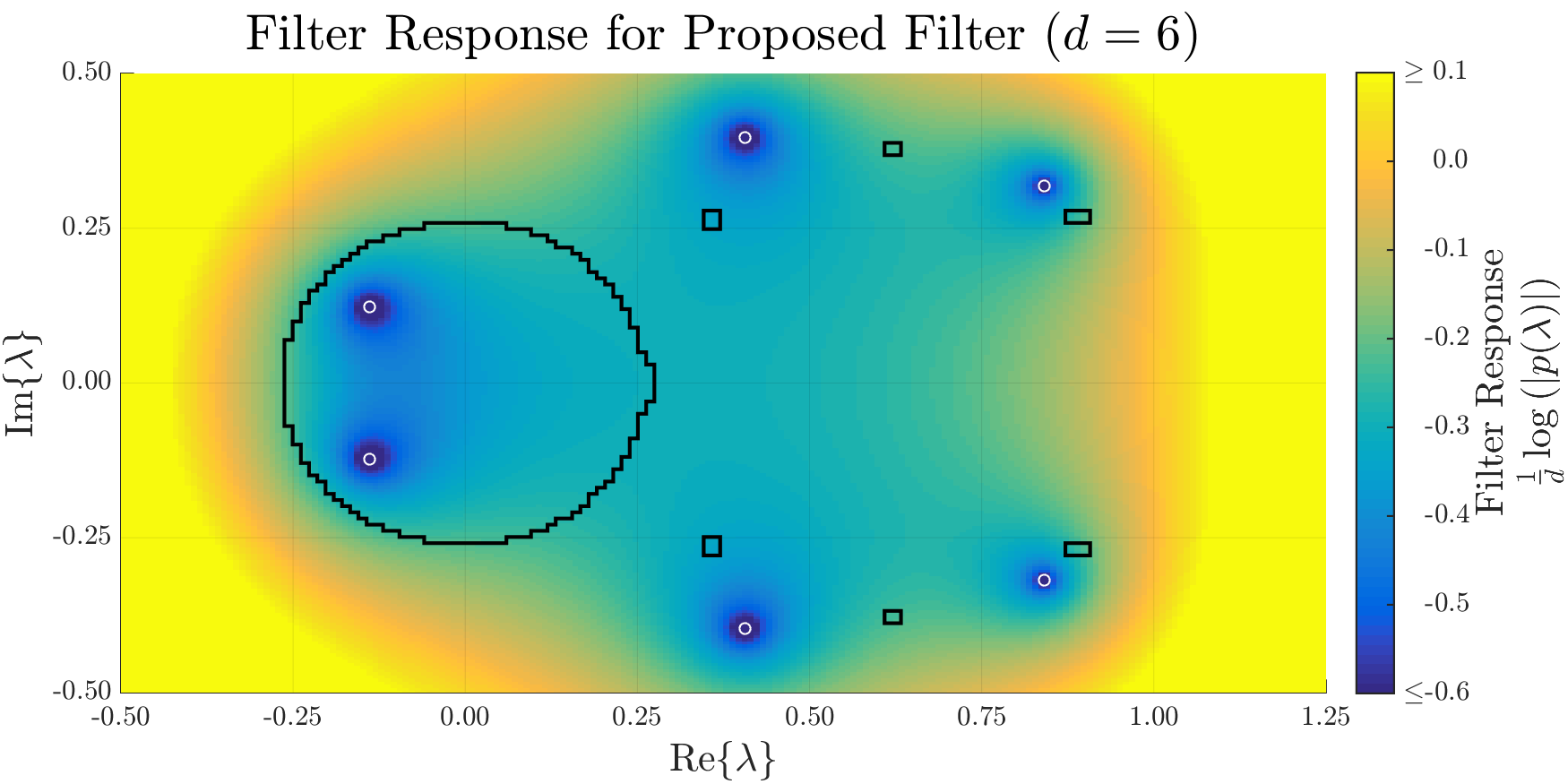}}{\caption{Proposed Filter ($d=6$)}\label{Sim2d3}}
\ffigbox[.220\textwidth]{\includegraphics[width=\linewidth]{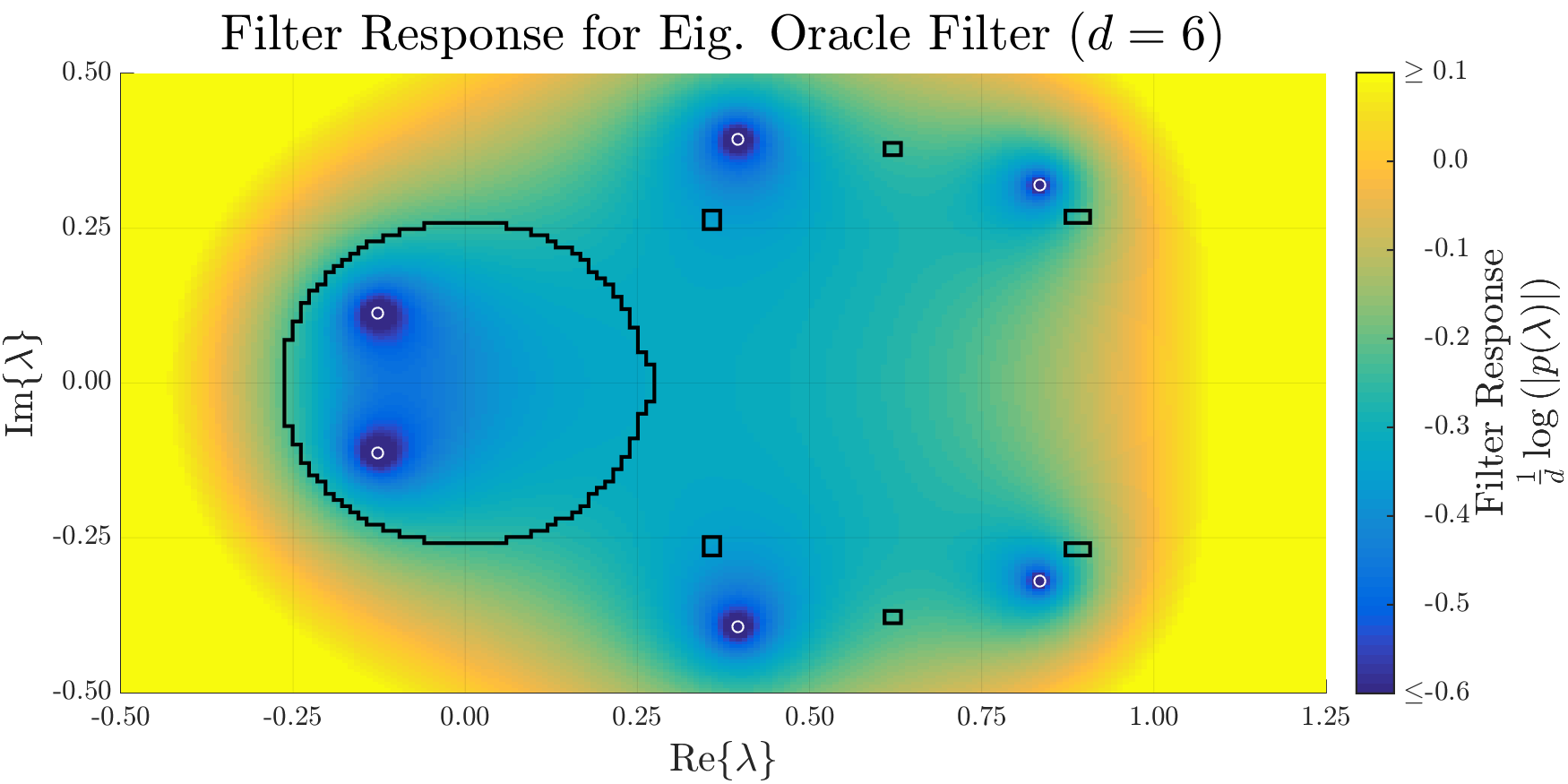}}{\caption{Ex. Optimal Filter ($d=6$)}\label{Sim2d4}}
\vspace{-3pt}
\end{subfloatrow}

%\begin{minipage}{.475\textwidth}\centering
%%\ffigbox[\textwidth]{
%\begin{minipage}{\textwidth}\centering
%\begin{subfloatrow}[2]
%\ffigbox[.475\textwidth]{\includegraphics[width=\linewidth]{NumSim2D1}}{\caption{Trivial Filter ($d=3$)}\label{Sim2d1}}
%\ffigbox[.475\textwidth]{\includegraphics[width=\linewidth]{NumSim2D2}}{\caption{Mean Matrix SDP ($d=3$)}\label{Sim2d2}}
%\end{subfloatrow}\vspace{5pt}
%\begin{subfloatrow}[2]
%\ffigbox[.475\textwidth]{\includegraphics[width=\linewidth]{NumSim2D3}}{\caption{Proposed Filter ($d=3$)}\label{Sim2d3}}
%\ffigbox[.475\textwidth]{\includegraphics[width=\linewidth]{NumSim2D4}}{\caption{Ex. Optimal Filter ($d=3$)}\label{Sim2d4}}
%\end{subfloatrow}
%\end{minipage}
%%}{\caption{}\label{Sim1d}}
%\end{minipage}
%\hfill
%\begin{minipage}{.475\textwidth}\centering
%%\ffigbox[\textwidth]{
%\begin{minipage}{\textwidth}\centering
%\begin{subfloatrow}[2]
%\ffigbox[.475\textwidth]{\includegraphics[width=\linewidth]{NumSim2D5}}{\caption{Trivial Filter ($d=6$)}\label{Sim2d5}}
%\ffigbox[.475\textwidth]{\includegraphics[width=\linewidth]{NumSim2D6}}{\caption{Mean Matrix SDP ($d=6$)}\label{Sim2d6}}
%\end{subfloatrow}\vspace{5pt}
%\begin{subfloatrow}[2]
%\ffigbox[.475\textwidth]{\includegraphics[width=\linewidth]{NumSim2D7}}{\caption{Proposed Filter ($d=6$)}\label{Sim2d7}}
%\ffigbox[.475\textwidth]{\includegraphics[width=\linewidth]{NumSim2D8}}{\caption{Ex. Optimal Filter ($d=6$)}\label{Sim2d8}}
%\end{subfloatrow}
%\end{minipage}
%%}{\caption{}\label{Sim1e}}
%\end{minipage}
\end{minipage}
}{\caption{Filter response magnitudes (log scale, per degree) for each filter type plotted in Fig.~\ref{Sim2c} (trivial filter, mean matrix SDP, proposed filter based on approx. density, and an example filter designed with known eigenvalues) are shown for degree $d=6$.  Locations of zeros are marked (white circles), and the boundary of the region $\Lambda_{\kappa,\tau}$ is shown (black contour).\vspace{-15pt}}\label{Sim2d}}

\end{floatrow}

\end{figure*}

\newcommand{\scval}{.85}
\newcommand{\scvalsm}{.7}
\newcommand{\scvalsmsm}{.6}

The empirical spectral distribution and empirical spectral density for a non-Hermitian matrix $\Xi_N$ with eigenvalues $\lambda_i\left(\Xi_N\right)$ are respectively given by the following functions.
\begingroup
\thinmuskip=.1\thinmuskip
\medmuskip=.1\medmuskip
\thickmuskip=.1\thickmuskip
\begin{align}
\kern-.5em
F_{\Xi_N}\left(x,y\right)&=\scalebox{\scval}{$\displaystyle\frac{1}{N}\sum_{i=1}^{i=N}$}\chi\left(x\leq\Re{\lambda_i\left(\Xi_N\right)},y\leq\Im{\lambda_i\left(\Xi_N\right)}\right) \kern-.1em\\
\kern-.5em f_{\Xi_N}\left(x,y\right)&=\scalebox{\scval}{$\displaystyle\frac{1}{N}\sum_{i=1}^{i=N}$}\delta(x-\Re{\lambda_i\left(\Xi_N\right)},y-\Im{\lambda_i\left(\Xi_N\right)})\kern-.1em
\end{align}
\endgroup
While these functions inherit the randomness of the eigenvalues of $\Xi_N$, their limiting behavior can sometimes be analyzed for useful information.  The following theorem of Girko~\cite{VGir1} allows analysis of random non-Hermitian matrices with independent entries.  Because the theorem accommodates non-identically distributed entries, it can be used to describe the adjacency matrices of random directed percolation networks with independent links.
\begin{theorem}[Girko's K25 Equation, abr.~\cite{VGir1}]\label{GirkoK25} 
\hfill\mbox{Let $\Xi_N$ be a} family of complex-valued $N\times N$ random matrices with independent entries that satisfy several regularity conditions. (See Theorem 25.1 of~\cite{VGir1} for the full list.)  Let $\Xi_N$ have expectation $B_N=\Exp{\Xi_N}$ and centralization $H_N=\Xi_N-B_N$ with entry variance $\sigma_{N,ij}^2=\operatorname{E}[|\left(H_N\right)_{ij}|^2]$.  Then 
\begin{equation}
\lim_{\beta\rightarrow 0^+} { \lim_{N\rightarrow \infty} {\left\|F_{\Xi_N}\left(x,y\right)-\widehat{F}_{\Xi_N,\beta}\left(x,y\right)\right\|}}=0
\end{equation}
almost surely, where
\begingroup
\thinmuskip=.1\thinmuskip
\medmuskip=.1\medmuskip
%\thickmuskip=0mu
\begin{equation}
\kern-.6em\scalebox{.9}{$\displaystyle\frac{\partial^2\widehat{F}_{\Xi_N,\beta}\left(t,s\right)}{\partial x \partial y}
 =\left\{\kern-.8em\begin{array}{cc} -\kern-.1em\frac{1}{4\pi}\kern-.1em\int_\beta^\infty \kern-.3em \left(\kern-.3em\frac{\partial^2}{\partial t^2}\kern-.1em+\kern-.1em\frac{\partial^2}{\partial s^2}\kern-.3em\right)\kern-.1em m_N\left(u,t,s\right)du & \kern-.5em (t,s)\kern-.3em\notin\kern-.3em G \\ \!\!\!\!\! 0 & \kern-.5em \left(t,s\right) \kern-.3em\in\kern-.3em G \end{array}\right.\kern-.8em$} \label{GirkoK25_Dens}
\end{equation}
\endgroup
(with the region $G$ defined below) and
\begingroup
\thinmuskip=.05\thinmuskip
\medmuskip=.05\medmuskip
%\thickmuskip=0mu
\begin{equation}
\begin{aligned}
m&_N\left(u,t,s\right)\kern-.2em = \scalebox{\scval}{$\kern-.2em \frac{1}{N}\tr\left[\left(C_1\left(u,s,t\right)+\ldots \vphantom{\left(B-(t+\im s)I_N\right)C_2\left(s,t\right)^{-1}\left(B_N-(t+\im s)I\right)^{*}}\right.\right.$}  \\
&\kern-.2em \scalebox{\scval}{$\left.\vphantom{\frac{1}{N}}\left. \vphantom{(C_1\left(s,t\right)+}\left(B_N-(t+\im s)I\right)C_2\left(u,s,t\right)^{-1}\left(B_N-(t+\im s)I\right)^{*}\right)^{\kern-.2em -1}\right]$}
\end{aligned}
\end{equation}
\endgroup
for $u>0$.  The matrices $C_1\left(u,s,t\right)$ and $C_2\left(u,s,t\right)$ are diagonal matrices with entries that satisfy the system of equations
\begingroup
\allowdisplaybreaks
\thinmuskip=.1\thinmuskip
\medmuskip=.1\medmuskip
%\thickmuskip=0mu
\begin{align}
&\begin{aligned}\kern-1em
(C_1&)_{kk}\left(u,s,t\right)=u+\scalebox{\scval}{$\displaystyle\sum_{j=1}^{j=N}\sigma_{N,kj}^2\left[\left(C_2\left(u,s,t\right)+\ldots\vphantom{\left(B_N-(t+s\im)I\right)^*C_1\left(u,s,t\right)^{-1}\left(B_N-(t+s\im)I\right)}\right.^{\vphantom{-1}}\right.$} \\
&\scalebox{\scval}{$\left.\left.\vphantom{C_2\left(y,s,t\right)+}\left(B_N-(t+s\im)I\right)^*C_1\left(u,s,t\right)^{-1}\left(B_N-(t+s\im)I\right)\right)^{\kern-.2em -1}\right]_{\kern-.2em jj \kern-2em}$}
\end{aligned}\label{GirkoK25_Sys1}
\\
&\begin{aligned}\kern-1em
(C_2&)_{\ell\ell}\left(u,s,t\right)=1+\scalebox{\scval}{$\displaystyle\sum_{j=1}^{j=N}\sigma_{N,j\ell}^2\left[\left(C_1\left(u,s,t\right)+\ldots\vphantom{\left(B_N-(t+s\im)I\right)C_2\left(y,s,t\right)^{-1}\left(B_N-(t+s\im)I\right)^*} \right.^{\vphantom{-1}}\right.$} \\
&\scalebox{\scval}{$\left.\left.\vphantom{C_1\left(y,s,t\right)+}\left(B_N-(t+s\im)I\right)C_2\left(u,s,t\right)^{-1}\left(B_N-(t+s\im)I\right)^*\right)^{\kern-.2em -1}\right]_{\kern-.2em jj \kern-2em}$}
\end{aligned}\label{GirkoK25_Sys2}
\end{align}
\endgroup
for $k,\ell=1,\ldots N$.  There exists a unique solution to this system of equations among real positive analytic functions in $u>0$.  The region $G$ is given by
\begingroup
\thinmuskip=.1\thinmuskip
\medmuskip=.1\medmuskip
%\thickmuskip=0mu
\begin{equation}
G=\scalebox{\scval}{$\left\{ (t,s) \middle|\displaystyle \limsup_{\beta\rightarrow 0^+}\limsup_{N\rightarrow\infty}\left|\frac{\partial}{\partial \beta}m_N\left(\beta,t,s\right)\right|<\infty\right\}.$}
\end{equation}
\endgroup
\end{theorem}
%\newpage
Because the solution is unique, it can be found through an iterative fixed point search.  Random networks with adjacency matrices satisfying the conditions of Theorem~\ref{GirkoK25} but with no additional special properties can always be analyzed through brute force.  However, this can be quite impractical as this would involve iterating on $2N$ variables and doing numerous matrix inversions for very large $N$ for each required value of $(u,t,s)$.  The most helpful property for computation is that the distribution have a symmetry group (with respect to node permutations) that acts transitively on the node.  That is, for every pair of nodes there is a permutation taking the first to the second that preserves the network distribution.  Intuitively, this means there are no statistically distinguishable nodes in the random network distribution.  In practical terms, this implies that $C_1,C_2$ are scalar matrices, the variance matrix has all row and column sums equal, and the system of equations~\mbox{\eqref{GirkoK25_Sys1}-\eqref{GirkoK25_Sys2}} can be rewritten in the following form (with 
\begingroup
\thinmuskip=.01\thinmuskip
\medmuskip=.01\medmuskip
\thickmuskip=.01\thickmuskip
$z=t+si$
\endgroup
) by applying the trace function to each half of~\mbox{\eqref{GirkoK25_Sys1}-\eqref{GirkoK25_Sys2}}.
\begingroup
\thinmuskip=.01\thinmuskip
\medmuskip=.01\medmuskip
\thickmuskip=.01\thickmuskip
\begin{align}
\kern-1emc_1&=u+\left(\scalebox{\scvalsmsm}{$\displaystyle\frac{1}{N}{\sum_{k=1}^{k=N}}$}\sigma_{N,kj}^2\right)
\scalebox{\scvalsmsm}{$\displaystyle\sum_{r=1}^{r=N}$}
\left({c_2+1/c_1
\scalebox{\scvalsm}{$ \lambda_r\left(\left(B_N-zI\right)^*\left(B_N-zI\right)\right)$}}\right)^{-1}\kern-1em\\
\kern-1emc_2&=1+\left(\scalebox{\scvalsmsm}{$\displaystyle\frac{1}{N}{\sum_{\ell=1}^{\ell=N}}$}\sigma_{N,j\ell}^2\right)
\scalebox{\scvalsmsm}{$\displaystyle\sum_{r=1}^{r=N}$}
\left({c_1+1/c_2 
\scalebox{\scvalsm}{$ \lambda_r\left(\left(B_N-zI\right)\left(B_N-zI\right)^*\right)$}}\right)^{-1}\kern-1em
\end{align}  
\endgroup

For the general case of random matrices that have different distribution than their transpose, this can still be a computationally difficult numerical problem because it requires computing the eigenvalues of 
\begingroup
\thinmuskip=.01\thinmuskip
\medmuskip=.01\medmuskip
%\thickmuskip=0mu
$\left(B_N-zI\right)\left(B_N-zI\right)^*$
\endgroup
, a problem which scales severely with $N$, for each required 
\begingroup
\thinmuskip=.01\thinmuskip
\medmuskip=.01\medmuskip
\thickmuskip=.01mu
$z=t+is.$
\endgroup
However, with random matrix distributions for which the mean $B_N$ is a normal matrix, $B_N-z I$ is normal so these eigenvalues are~\cite{GAll1}
\begingroup
\thinmuskip=.1\thinmuskip
\medmuskip=.1\medmuskip
%\thickmuskip=0mu
\begin{equation}
\lambda_r\left(\left(B_N-zI\right)\left(B_N-zI)^*\right)\right)=\left|\lambda_r\left(B_N\right)-z\right|^2.
\end{equation}
\endgroup
Therefore, the number of eigenvalue computations can be reduced to one for normal $B_N$.  Hence, distributions for which the mean matrix is normal can be solved with no more computational burden than in the symmetric mean case (actually a subcase of normal matrices), which was examined in~\cite{SKru6}.

\begin{remark}[Numerical Integration Steps]
\hfill \mbox{Solving for the} density $f_{\Xi_N,\beta}$ from the $m_N(u,t,s)$ function via~\eqref{GirkoK25_Dens} requires numerical integration with respect to $u$ from a  small value ($\beta=10^{-6}$ used in simulations) to a large upper limit ($10^2$ used in simulations).  Furthermore, the integration occurs in a region (compliment of $G$) where $\frac{\partial}{\partial u}m_N(u,t,s)$ approaches infinite magnitude for small $u$.  Therefore, logarithmically spaced integration interval endpoints are recommended.
\end{remark}

\begin{example}[Asymmetric Stochastic Block Model]\hfill
\mbox{For the} consensus application simulation in Section~\ref{NumericalSimulations}, a spectral density approximation is required for a transpose-asymmetric stochastic block model network with iteration matrix $W=I-\alpha \widehat{L}_R$ where $\widehat{L}_R$ is the directed, row-normalized Laplacian.  Under suitable conditions, this can be \mbox{accomplished by}
\begin{equation}
\widehat{f}_{W_N,\beta}(x,y)=\scalebox{\scval}{$\frac{1}{\alpha^2}$}\widehat{f}_{\Xi_N,\beta}\left(\scalebox{\scval}{$\frac{x-1}{\alpha}+1,\frac{y}{\alpha}$}\right)
\end{equation}
 where $\Xi_N=\frac{1}{\gamma}A_N$ and $\gamma$ is the expected row sum.  In a directed stochastic block models with $N=MS$ nodes divided among $M$ populations of $S$ nodes each, a node in population $i$ forms a links to each nodes in population $j$ independently with probability $\Theta_{ij}$ depending on the two populations. If the $M\times M$ matrix $\Theta$ is a normal matrix that is invariant under equal row and column permutations that act transitively on the populations, then the mean adjacency matrix $B_N=\Theta \otimes \mathbf{1}_{S\times S}-\Theta_{11} I$ is normal by Kronecker product of normal matrices.  In Section~\ref{NumericalSimulations} this is applied to a stochastic block model with an inter-population structure described by a directed cycle.
\end{example}

%\newpage
\vspace{0pt}
\section{Directed Networks:~~Consensus Filter Application and Numerical Simulations}\label{NumericalSimulations}
%\vspace{2pt}
%\input{Sections/FilterDesign_ComplexDN_NumericalSimulations}
An application that benefits from good spectral density approximations, consensus acceleration filters apply a filter to the consensus state at each node to achieve faster convergence.  Several example consensus filter design methods using this can be found in~\cite{SSun1,ASan4,EKok1,EMon1,ALou1,SApe1,FGam1}.  One such approach periodically applies a filter to the consensus state every $d$ iterations, where $d$ is the filter degree.  Examples that formulate optimization problems for filter design using the spectral asymptotics of large-scale random graphs include~\cite{SKru1,SKru2,SKru3,SKru4,SKru5,SKru6}.

In particular,~\cite{SKru6} uses Girko's K25 method for directed random network models with transpose-symmetry.  It then proposes the following optimization problem for non-time-varying random networks to approximately optimize the convergence rate $\frac{1}{d}\ln\rho\left(p(W)-J_{\boldsymbol\ell}\right)$.  Here $W$ is the consensus iteration matrix, ${\boldsymbol\ell}$ is the left eigenvector of $W$ corresponding to eigenvalue $\lambda=1$, $J_{\boldsymbol\ell}=\mathbf{1}{\boldsymbol\ell}^\top/{\boldsymbol\ell}^\top\mathbf{1}$ is the $\boldsymbol\ell$-weighted average consensus transform, $\rho$ is the spectral radius, and polynomial $p$ describes the filter coefficients $\{a_k\}_{k=1}^{k=d}$.  Sample points $\Lambda_S$ are chosen within the identified filtering regions $\Lambda_{\kappa,\tau}$ derived from the approximate spectral density $f_{W_N,\beta}$ (small values of $\beta,\kappa,\tau)$.  The filter response is minimized at these points through the following QCLP.
\begingroup
\thinmuskip=.1\thinmuskip
\medmuskip=.1\medmuskip
\thickmuskip=.1\thickmuskip
\allowdisplaybreaks
\begin{equation}\label{Problem3}
\begin{gathered}
\kern-.4em\begin{aligned}
\smash{\min_{\mathbf{a}\in \mathbb{R}^{d+1}\!,\varepsilon}} \enskip \varepsilon \qquad
\mrlap{\st}\hphantom{\min_{\mathbf{a}}} \enskip & \mathbf{1}^\top\mathbf{a}=1 \\
&\mathbf{a}^\top Q\left(\lambda_{i}\right)\mathbf{a} < \varepsilon
\quad \textrm{for all}~ \lambda_i\in \Lambda_S
%\\
%&~\hphantom{\mathbf{a}^\top Q\left(\lambda_{i}\right)\mathbf{a}<\varepsilon}\mllap{\textrm{for all}~ \lambda_i\in \Lambda_S}
\end{aligned}\\
\Lambda_S\subseteq\Lambda_{\kappa,\tau}=\left\{\left|\lambda-1\right|>\kappa \middle| \widehat{f}_{W_N,\beta}\left(\Re{\lambda},\Im{\lambda}\right)>\tau\right\}
\end{gathered}
\end{equation}
\endgroup
where $Q\left(\lambda_i\right)$ is the real, positive semidefinite matrix
\begin{gather}
Q\left(\lambda_i\right)=\scalebox{1}{$\displaystyle\frac{1}{2}$}\left(V\left(\lambda_i\right)^*V\left(\lambda_i\right)+V\left(\overline{\lambda}_i\right)^*V\left(\overline{\lambda}_i\right)\right)\\
V\left(\lambda_i\right)=\left[\lambda_i^0,\ldots,\lambda_i^d\right].
\end{gather}
This section displays numerical simulation results demonstrating approximate spectral densities found for consensus iteration matrices (row-normalized directed Laplacian weights $W=I-\alpha \widehat{L}_R$ used) of transpose-asymmetric directed network models via the computational simplifications justified by Section~\ref{NonSymmetricMean}.  The simulations also design consensus filters and evaluate their comparative performances.

%~\\ \vspace{-5pt}

For a stochastic block model with $M=5$ populations of size $M=200$ arranged in a cycle and link probability $\Theta_{ii}=.05$ within the populations and $\Theta_{ij}=.03$ for $j\equiv i+1 (\textrm{mod}~N)$ to the next population, numerical simulations are shown.  Figure~\ref{Sim1a} shows the expected empirical spectral distribution of the consensus iteration matrix ($\alpha=1$) simulated over 1000 Monte-Carlo trials.  Figure~\ref{Sim1b} shows the deterministic approximation derived via Girko's K25 equation. Figure~\ref{Sim1c} compares the convergence rate per iteration of the filtered consensus process for the trivial filter (red), the filter with response optimized at the mean  iteration matrix eigenvalues (black, mean SDP method from~\cite{EKok1}), the filter designed according to the introduced optimization (blue), and  the filter designed after the network is drawn from the distribution (green).  Vertical histograms show the distribution of the trials for each filter type.  However, the spreads of the trial result distributions are small due to the limiting spectral behavior.   Figure~\ref{Sim1d} shows filter response plots for the various filter types. Similarly, for a stochastic block model with $S=8$ populations each of size $M=200$ and the same link probabilities as before, Figures \ref{Sim2a}-\ref{Sim2d} show analogous results.

Note that the proposed filter design performs nearly as well as the filter designed with pre-knowledge of the eigenvalues, although the existence of underperforming outliers is also observed.  Also note that this simulation demonstrates that knowing the eigenvalues of the mean iteration matrix is insufficient for filter design for these types of networks.  The notion of spread provided by the Girko approximation to the spectral density addresses this.

%\newpage
\section{Conclusion}\label{Conclusion}
This paper examined the spectral asymptotics of large-scale, random, non-symmetric matrix models, extending previous results to analyze models with distribution different from the distribution of their transpose.  Girko's K25 stochastic canonical equation method provides a valuable tool to analyze such matrices when they arise from a directed percolation model.  Under the condition that the large-scale random network model has a symmetry group that is transitive with respect to action on the nodes and the mean adjacency matrix is a normal matrix,  spectral density approximation via Girko's equations can be handled at no greater computation cost than is incurred for large-scale directed network models with equal link probability in \mbox{each direction.}

Therefore, if a large-scale random directed network distribution satisfies the properties, the described method can be applied, for instance, to spectral density approximation for the associated consensus iteration matrices.  The resulting deterministic approximate spectral density can then be used to inform consensus filter design optimization problems for convergence acceleration.  Included results from numerical simulations demonstrate good approximation quality for the empirical spectral distribution of large-scale random networks with link model transpose-asymmetry and also strongly support application of these approximate densities to consensus acceleration filter design via the method introduced in~\cite{SKru6}.  Continuing research efforts on this topic include analysis of filter design methods for time-varying random networks directed and analysis of network models with correlations among the directed link random variables.

\clearpage
\bibliographystyle{IEEEtran}
%\begin{spacing}{}
\begingroup
\small
%\bibliography{Sections/FilterDesign_ComplexDN_Bibliography}
\bibliography{FilterDesign_ComplexDN_Bibliography}
\endgroup
%\end{spacing}

\end{document}